\documentstyle[aps,twocolumn]{revtex}

\begin{document}

\input epsf
\newcommand{\infig}[2]{\begin{center}\mbox{ \epsfxsize #1
                       \epsfbox{#2}}\end{center}}

\newcommand{\be}{\begin{equation}}
\newcommand{\ee}{\end{equation}}
\newcommand{\bea}{\begin{eqnarray}}
\newcommand{\eea}{\end{eqnarray}}
\newcommand{\wee}[2]{\mbox{$\frac{#1}{#2}$}}   
\newcommand{\unit}[1]{\,\mbox{#1}}
\newcommand{\degree}{\mbox{$^{\circ}$}}
\newcommand{\ltish}{\raisebox{-0.4ex}{$\,\stackrel{<}{\scriptstyle\sim}$}}
\newcommand{\vs}{{\em vs\/}}
\newcommand{\bin}[2]{\left(\begin{array}{c} #1 \\ #2\end{array}\right)}
                        
\draft

\title{Dissipative dynamics of Bose condensates in optical cavities}

\author{Peter Horak and Helmut Ritsch}
\address{Institut f\"ur Theoretische Physik,
Universit\"at Innsbruck, Technikerstr.\ 25, A-6020 Innsbruck, Austria}
\date{\today; submitted to Phys.\ Rev.\ A}

\maketitle

\begin{abstract}
We study the zero temperature dynamics of Bose-Einstein condensates in driven
high-quality optical cavities in the limit of large atom-field detuning. We
calculate the stationary ground state and the spectrum of coupled atom and
field mode excitations for standing wave cavities as well as for travelling
wave cavities. Finite cavity response times lead to damping or controlled
amplification of these excitations. Analytic solutions in the Lamb-Dicke
expansion are in good agreement with numerical results for the full problem and
show that oscillation frequencies and the corresponding damping rates are
qualitatively different for the two cases. 
\end{abstract}


\narrowtext


\section{Introduction}

The experimental realization of Bose-Einstein condensation (BEC) in dilute
atomic gases a couple of years ago \cite{bec0} as a consequence of  improved
cooling and trapping techniques has dramatically boosted the study of ultracold
atoms.  Today, BEC is a widespread tool and a huge range of new phenomena has
been investigated experimentally and theoretically, see e.g.\
Refs.~\cite{Ketterle,Walls,Stringari} for recent overviews. In this context the
interaction of BECs with laser light and optical lattices has been studied
intensively  \cite{Phillips,Kasevich,Choi,Jaksch,Berg,Marzlin} and effects such
as the reduction of the speed of light by many orders of magnitude \cite{Hau}
and the occurrence of superradiance \cite{superrad} have been found. Recently
the optical creation of vortices \cite{Cornell,Dalibard} has been demonstrated
and many more intriguing effects have been theoretically predicted.

In parallel, the field of cavity quantum electrodynamics, which studies the
interaction of matter with one or a few single light modes, has reached such a
level of sophistication that the interaction of light with the internal and
external degrees of freedom of a single neutral particle can be observed and
controlled in a very precise way \cite{Haroche}. For optical fields trapping and
cooling of a single atom in a cavity mode has been demonstrated
\cite{Rempe,Kimble}. It is thus a logical next step to combine these two
successful techniques and study the interaction of a BEC with a high finesse
optical cavity, that is, the strong coupling of a far detuned optical mode to
the dynamics of a condensate described by its macroscopically occupied wave
function \cite{Milburn}.  Several groups have already been working along these
lines and, for example, predicted the amplification of matter waves \cite{Law}
and the occurrence of dressed condensates \cite{Meystre}. In the extreme limit
one could envisage a large atomic cloud trapped and manipulated with a single
photon.

In this work we extent our recently proposed scheme for cooling one or a few
atoms in high-quality optical cavities \cite{cavcool1,cavcool2} to the case of
a BEC. This requires a  quantum mechanical treatment of the external degrees of
freedom and the inclusion of atom-atom interactions.  The system under
investigation is a Bose-Einstein condensate interacting with the mode (the two
modes) of a driven standing wave cavity (ring cavity). The properties of such a
system as a measuring device for condensates have been discussed previously
\cite{Milburn,PRA}.

We will investigate the ground state and collective excitations 
\cite{Stringari2,Perez,You,Castin} of the coupled condensate-light field system
in the optical potential of the cavity. Because of the strong coupling of the
condensate wave function to the cavity modes, an oscillation of the condensate
also leads to an oscillation of the intracavity light fields. Accordingly, the
oscillation frequencies of the collective excitations are shifted with respect
to an external optical potential of fixed depth, i.e., the optical potential
formed by a free space standing wave. Furthermore, for appropriately chosen
parameters the dissipative dynamics of the cavity due to cavity decay gives
rise to damping of the condensate excitations without incoherent spontaneous
emission.  We analyze this damping mechanism and study its parameter dependence
by numerical solutions of the coupled equations of motion as well as by
analytic solutions of a simplified model based on the Lamb-Dicke expansion.

This paper is organized as follows. In Sec.~II we discuss the case of a
condensate interacting with the single mode of a standing wave cavity. After
presenting the set of coupled non-linear equations of motion for the condensate
and the light field, we discuss the numerical and analytical solutions for the 
ground state and the collective excitations. Section~III investigates the more
complicated situation of a condensate coupled to the two independent modes of
an optical ring cavity. In Sec.~IV we discuss the influence of binary
collisions between the atoms within the condensate on the excitation
frequencies and damping rates. Finally, we summarize our results in Sec.~V.


\section{BEC in standing wave cavity}

Let us first consider the case of a Bose-Einstein condensate interacting with a
single standing wave mode. The cavity mode is assumed for all times to be in a
coherent state $|\alpha(t)\rangle$ and the condensate is described by a single
wave function $|\psi(t)\rangle$ for all $N$ particles, which is a good
approximation at zero temperature.  This means that we factorize the quantum
state of the system and thus neglect any entanglement between the condensate
and the cavity field  which might build up in the course of the time evolution.
This simplification is only justified in the limit of a large photon number
$|\alpha|^2$.  To avoid spontaneous emission, we assume a very large detuning
of the light field from the atomic resonance. More precisely, we assume that
the cavity decay is the dominant incoherent process in the system and we thus
require that $\kappa\gg N\Gamma s$, where $\kappa$ is the cavity decay rate,
$N$ the number of atoms in the condensate, $\Gamma$ the spontaneous decay rate
of the atoms, and $s$ the atomic saturation parameter. After adiabatic
elimination of the atomic excited states we obtain the following equations of
motion,
\begin{mathletters}
\label{eq:st}
\bea
& & \frac{d}{dt} \alpha(t)  = 
   \left[ i\Delta_c - i N\langle U(\hat{x})\rangle -\kappa \right]
   \alpha(t) + \eta,  \label{eq:ast} \\
& & i \frac{d}{dt} \psi(x,t) = \nonumber \\
& & \quad\left\{ \frac{\hat{p}^2}{2m} + |\alpha(t)|^2 U(x) + 
          Ng_{coll}|\psi(x,t)|^2\right\} \psi(x,t). \label{eq:GPEst}
\eea
\end{mathletters}
Here $\Delta_c$ is the detuning of the pump field from the cavity mode, 
$U(x)=U_0 \cos^2(kx)$ is the optical
potential formed by a single cavity photon, and $\eta$ describes the action of
the driving laser. The expectation value $\langle U(\hat{x})\rangle$ has to be
taken with respect to the momentary wave function $|\psi(t)\rangle$.
Equation (\ref{eq:GPEst}) is the well known Gross-Pitaevskii
equation (GPE) for a condensate in an external field which in our case depends 
on the momentary field intensity $|\alpha(t)|^2$. The last term
in the GPE models the interaction of atoms within the condensate where 
$g_{coll}$ is related to the s-wave scattering length $a$ by 
$g_{coll} = 4\pi\hbar^2 a/m$. 

Equations (\ref{eq:st}) are two coupled nonlinear equations describing the
dynamics of the compound system formed by the condensate and the cavity field
\cite{PRA}. The most interesting effects occur for parameters where the
coupling between these equations significantly changes the system behaviour. We
thus impose the condition $N U_0\geq \kappa$, which guarantees that the presence
of the condensate shifts the cavity frequency efficiently into or out of
resonance with the driving field. At the same time the optical potential depth 
$|\alpha|^2 U_0$ should be large enough to provide at least a few bound states
for the atoms. Limitations and the interesting parameter regimes for this model
have been discussed in Ref.~\cite{PRA}.

\subsection{Ground state}

In order to obtain the ground state of the compound condensate-cavity system,
we have to find the stationary solution of the system of coupled nonlinear
equations (\ref{eq:st}). 
This can be done by elimination of $|\alpha(t)|^2$ in 
(\ref{eq:GPEst}) using (\ref{eq:ast}), and a subsequent numerical solution of
the resulting non-linear equation for the ground state wave function with the
method of steepest descent. This consists of a numerical propagation of the GPE
in imaginary time $\tau=it$ until the wave function converges to a stationary
state.

In this work we will concentrate on the case of $U_0>0$ where the potential
minima coincide with the antinodes of the field (low-field seeking atoms). The
ground state wave function will thus be localized at the field antinodes,
thereby minimizing the coupling of the condensate to the light field. For a
cavity resonant with the driving field, this means that the
photon number is maximum for the stationary ground state. Any
excitation of the condensate will then lead to a smaller cavity field.

As expected we find that the ground state wave function becomes better localized
for stronger driving fields $\eta$ and larger optical potentials $U_0$. On the
other hand, a strong atom-atom repulsion (large positive values of $g_{coll}$)
increases the width of the BEC wave function and thus counteracts the confining
effect of the potential. This in due course leads to an increased coupling of
the BEC to the cavity field and hence a smaller field intensity. A more detailed
analysis of the ground state wave function $\psi_0(x)$,
its energy $\mu$, and the stationary field intensity $|\alpha_0|^2$ has been
given in Ref.~\cite{PRA}.

\subsection{Collective excitations}

Let us now turn to weak excitations of the condensate from the ground state.
First, we will calculate the spectrum of collective excitations of the
condensate. In contrast to fixed external fields, the trapping potential in the
cavity depends on the BEC wave function. Hence, excitations include small
deviations of the wave function {\em and} the cavity field $\alpha$ from their
respective stationary state. We may thus write
$\psi(x,t)=\exp(-i \mu t)[\psi_0(x)+\delta\psi(x,t)]$ and 
$\alpha(t) = \alpha_0+\delta\alpha(t)$. For convenience we have already 
included the ground state time evolution into the ansatz for the wave function
here. Inserting this into Eqs.~(\ref{eq:st}) and linearizing in $\delta\psi$
and $\delta\alpha$ we obtain
\begin{mathletters}
\label{eq:stlin}
\bea
i \frac{d}{dt}\delta\alpha & = & 
   [-\Delta_c+N\langle \psi_0|U(\hat{x})|\psi_0\rangle -i\kappa]\delta\alpha 
   \nonumber \\
& &   + N\alpha_0 \langle\delta\psi|U(\hat{x})|\psi_0\rangle
   + N\alpha_0 \langle\psi_0|U(\hat{x})|\delta\psi\rangle, 
   \label{eq:stlina} \\
i \frac{d}{dt}\delta\psi & = & 
   \left\{ \frac{\hat{p}^2}{2m} + |\alpha_0|^2 U + 
      2Ng_{coll}|\psi_0|^2 - \mu\right\} \delta\psi \nonumber \\
& &   + Ng_{coll}\psi_0^2 \delta\psi^*
   + \alpha_0 U \psi_0 \delta\alpha^* + \alpha_0^* U \psi_0 \delta\alpha.
\eea
\end{mathletters}
For large $\kappa$ (more precisely, for $1/\kappa$ much smaller than the 
time scale of the condensate motion), the cavity field follows adiabatically the
changes of the wave function and thus $\delta\alpha$ can be adiabatically
eliminated. In this case one recovers the limit of Ref.~\cite{PRA}.

In general the linearized time evolution couples
the deviations $\delta\psi$ and $\delta\alpha$ also to their complex conjugates.
In order to obtain excitation eigenstates, i.e., periodic solutions, we 
thus have to use the simultaneous ansatz
\bea
\delta\psi & = & e^{-\gamma t} \left[ 
   e^{-i\nu t} \delta\psi_+(x) + e^{i\nu t} \delta\psi_-(x)^* 
   \right], \nonumber\\
\delta\alpha & = & e^{-\gamma t} \left[ 
   e^{-i\nu t} \delta\alpha_+ + e^{i\nu t} \delta\alpha_-^*\right].
\eea
The collective excitations are thus defined as the solutions of the eigenvalue
problem
\be
\omega \left( \begin{array}{c} 
      \delta\alpha_+ \\ \delta\alpha_- \\ \delta\psi_+(x) \\ \delta\psi_-(x)
   \end{array}\right)
= {\bf M} \left( \begin{array}{c} 
      \delta\alpha_+ \\ \delta\alpha_- \\ \delta\psi_+(x) \\ \delta\psi_-(x)
   \end{array}\right),
\label{eq:eigst}
\ee
where ${\bf M}$ is easily obtained from Eqs.\ (\ref{eq:stlin}) as a
non-Hermitian matrix. The complex eigenvalues have the form 
$\omega_n = \nu_n -i \gamma_n$, where $\nu_n$ is the oscillation frequency of
the $n$th collective excitation and $\gamma_n$ the corresponding damping rate.
Note that, depending on the parameters, negative damping rates are possible,
leading to an exponential growth of the collective excitations. In this case
the assumption of small deviations from the ground state imposed above only
holds for very short times.  Hence by changing some cavity parameters we can
switch between stable and unstable cases and generate controlled excitations of
the condensate and study their decay. In the following we will, however,
concentrate on the case of positive $\gamma_n$ and therefore damped
excitations.

Physically this damping arises from a kind of Sisyphus mechanism. For cavity
damping rates $\kappa$ of the order of the oscillation frequencies $\nu_n$, the
cavity field follows with a certain delay the changes of the condensate
wave function. By properly choosing the system parameters, it can be achieved
that on average the wave function has to climb up the potential hills at higher
cavity field intensities and runs down at lower intensities. The condensate
thus loses potential energy which is carried away by the cavity output field
without an intrinsic decoherence of the condensate.

Furthermore it should be emphasized that the appearance of a damping rate in
the linearized equations (\ref{eq:stlin}) is a purely quantum feature related
to the width of the atomic wave function. In the semiclassical limit of a
point-like particle, the self-consistent ground state yields a particle exactly
located at the antinodes of the cavity and hence all expectation values in 
Eq.~(\ref{eq:stlina}) vanish. Thus the cavity field decouples from the atomic
degrees of freedom and no damping of the atomic motion occurs to lowest order 
in the elongation $x$. This is in contrast to the case of a ring cavity as will
be shown in the following section.

\begin{figure}[tb]
\infig{22em}{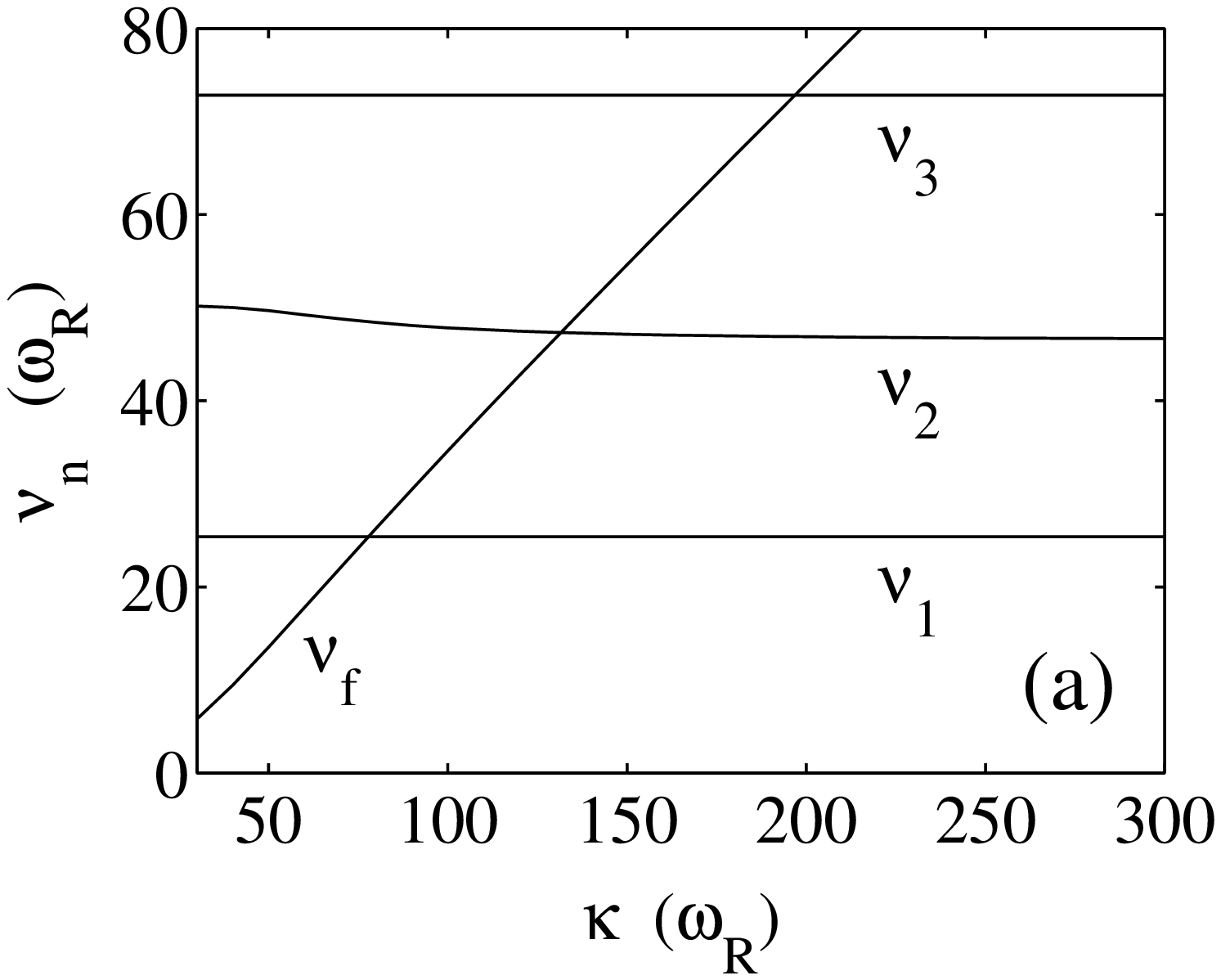}
\infig{22em}{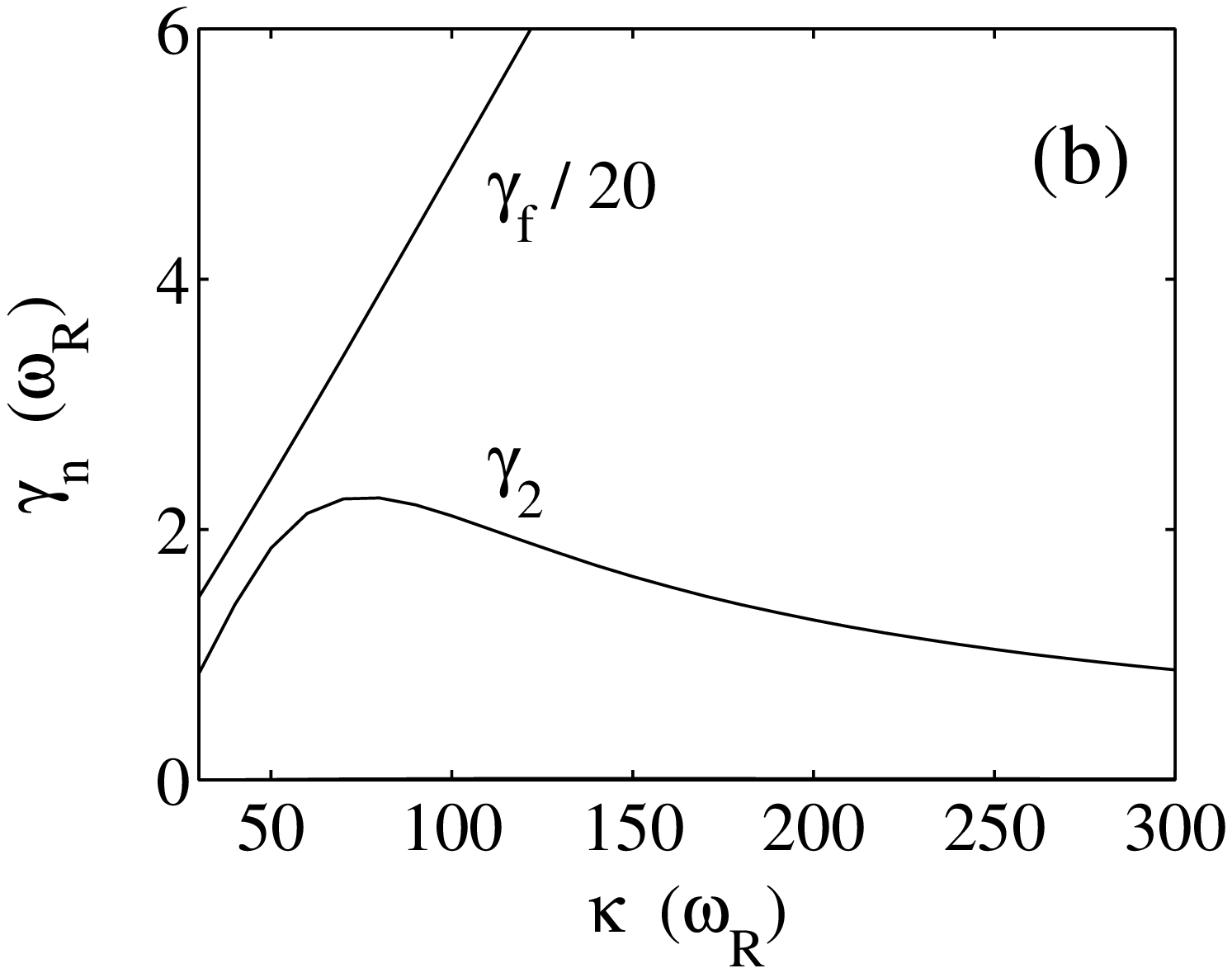}
\caption{Collective condensate excitations: (a) frequencies $\nu_n$ and (b) 
 decay rates $\gamma_n$ vs cavity decay rate $\kappa$. The parameters are
 $g_{coll}=\Delta_c=0$, $N U_0=10\kappa$, $\eta^2=20\kappa N \omega_R$.
}
\label{fig:A}
\end{figure}

In Fig.~\ref{fig:A} we show the oscillation frequencies and damping rates of the
lowest collective excitations obtained numerically by calculating the
eigenvalues of Eq.\ (\ref{eq:eigst}) on a spatial grid. The eigenvalues are
plotted as a function of the cavity decay rate $\kappa$. Note that in order to
keep the optical potential constant, we also have to scale the driving field
$\eta^2$ and the optical potential per photon $U_0$ proportional to $\kappa$.

We see that there exists one single eigenvalue $\omega_f=\nu_f-i\gamma_f$ 
which scales approximately
proportional to $\kappa$ in contrast to all of the other eigenvalues. This
specific excitation of the system corresponds to an eigenmode where mainly the
cavity field oscillates and the condensate wave function is only weakly
perturbed. In fact, equation (\ref{eq:stlina}) shows that in the case where the
atoms are well localized at the antinodes of the field (semiclassical limit),
the cavity mode decouples from the matter wave function and the eigenvalue is
given by $\omega_f = -\Delta_c -i\kappa$.

Second, we notice that out of the other modes $\nu_n$, $n=1,2,\dots$, the ones
with odd indices are independent of $\kappa$ and their damping rates vanish.
This effect is due to the spatial symmetry of the problem considered here. For
all parameters we find that the ground state wave function $\psi_0$ is symmetric
in the position $x$. Thus for all {\em antisymmetric} excitations the
expectation values in Eq.\ (\ref{eq:stlina}) vanish and the light field
decouples. Therefore these odd ($n=1,3,\dots$) excitations are the same as for a
trap of constant light intensity and hence there is no Sisyphus damping
mechanism at work. Consequently only the lowest {\em symmetric} collective
excitations are significantly altered by the interaction with the damped
cavity mode. We will discuss the
parameter dependence of the excitation $n=2$ by using an approximate analytic
solution in the next section.

Let us finally emphasize that the oscillation frequency and the damping rate of
the symmetric collective excitations can be monitored nondestructively 
via the cavity output intensity.

\subsection{Harmonic oscillator approximation}
\label{sec:stharm}

In order to gain more insight into the parameter dependence of this damping
mechanism, we will now analytically solve an approximate model of our system.
To this end we expand the optical potential $U(x)=U_0\cos^2(kx)$ with $U_0>0$
up to second order around the antinodes of the field, i.e., we set 
$U(x)=U_0 (kx)^2$ and assume $g_{coll}=0$, i.e., no atom-atom interaction. 
For simplicity we will also assume $\Delta_c=0$.

The ground state of the Schr\"odinger equation (\ref{eq:GPEst}) is thus the well
known harmonic oscillator ground state which depends on the cavity field
$|\alpha|^2$ in a parametric way. After inserting this wave function in the
expectation value in Eq.\ (\ref{eq:ast}) we obtain an equation for the 
self-consistent cavity field with the solution
\be
|\alpha_0|^2 = \frac{\eta^2}{\kappa^2} - 
    \frac{N^2 U_0 \omega_R}{4\kappa^2}.
\ee
The corresponding harmonic oscillator frequency is then
\be
\omega_0 = 2 \omega_R \sqrt{|\alpha_0|^2 U_0/\omega_R}
\ee
and the ground state energy is $\mu=\omega_0/2$.

For the collective excitations we now have to solve Eqs.\ (\ref{eq:stlin}) with
the harmonic potential. The last expectation value of Eq.\ (\ref{eq:stlina})
thus reads
\bea
\langle\psi_0|U(\hat{x})|\delta\psi\rangle & = & 
U_0 \langle\psi_0|(k\hat{x})^2|\delta\psi\rangle \nonumber \\
& = & -U_0 \frac{\omega_R}{\omega_0} 
  \langle\psi_0|(a-a^{\dagger})^2|\delta\psi\rangle,
\eea
where we have used the standard relation between the position operator $\hat{x}$
and the ladder operators $a$ and $a^{\dagger}$ of the harmonic oscillator. From
this we see that the cavity field only couples to wave function deviations
$\delta \psi$ containing the ground state $\psi_0$ and/or the second excited
state $\psi_2$ of the harmonic oscillator. Most of the harmonic oscillator
excited states are thus unperturbed and we find the collective excitations of
the form $\delta\psi_+(x)=\psi_n$, 
$\delta\alpha_+ = \delta\alpha_- = \delta\psi_-(x)=0$ with the positive
eigenvalues $\omega=n\omega_0$ for $n=1$ and $n\ge 3$. Analogously there exist
excitations with negative eigenvalues $\omega=-n\omega_0$ of the form
$\delta\psi_-(x)=\psi_n$, 
$\delta\alpha_+ = \delta\alpha_- = \delta\psi_+(x)=0$.
Hence, in addition to the
antisymmetric states already found to decouple previously, also the higher lying
symmetric states decouple in the harmonic approximation. Therefore these
symmetric excitations are only damped due to the anharmonicity of the potential
and due to atomic collisions in the full model.

The remaining (and most interesting) collective excitations are finally found
by restricting the wave functions $\delta\psi_{\pm}$ in Eq.\ (\ref{eq:eigst}) to
the 2-dimensional Hilbert space spanned by $\psi_0$ and $\psi_2$. The resulting
6x6 matrix has two zero eigenvalues and the other four eigenvalues have to be
found by solving the 4th order polynomial equation
\bea
& & (-i\kappa+NU_0\case{\omega_R}{\omega_0}-\omega)
(-i\kappa-NU_0\case{\omega_R}{\omega_0}-\omega)
(\omega^2-4\omega_0^2) \nonumber \\
& & -4(NU_0\omega_R)^2=0.
\label{eq:polst}
\eea
This gives us the (complex) eigenvalues $\omega_f$ and $\omega_2$ and their
counterparts of negative frequency.
Although an analytic solution of (\ref{eq:polst}) is possible in principle, the
resulting expressions are rather long and do not provide much insight. Instead,
we calculate the eigenvalue $\omega_2$ in the limit of large $\kappa$ as in
Fig.\ref{fig:A}, i.e., by keeping $U_0/\kappa$ and $\eta^2/\kappa$ constant. 
The zeroth order in this expansion in $\omega/\kappa$ yields the leading order
of the frequency
\be
\nu_2 = 2\omega_0\sqrt{1-\frac{N^2 U_0 \omega_R}{4\eta^2}} 
\label{eq:nu2st}
\ee
and the first order gives the leading order of the decay rate
\be
\gamma_2 = \frac{4 N^2 U_0^2 \omega_R^2}{\kappa^3}
   \left(1-\frac{N^2 U_0 \omega_R}{4\eta^2} \right)^2.
\label{eq:g2st}
\ee
Equation (\ref{eq:nu2st}) gives a quantitative explanation for the frequency
shift of $\nu_2$ according to the coupling of the BEC and the cavity mode as
compared to the value $2\omega_0$ for the case of a harmonic oscillator
potential of fixed photon number. We also see that the small variation of
$\nu_2$ in Fig.\ref{fig:A}(a) for small values of $\kappa$ are in fact of the
order $1/\kappa^2$. Equation (\ref{eq:g2st}) leads to the asymptotic behaviour
like $1/\kappa$ for the decay rate $\gamma_2$ in Fig.\ref{fig:A}(b). 
In the limit of a strong driving field the frequency $\nu_2$ 
of the second collective excitation approaches the harmonic oscillator value.
Simultaneously the damping rate $\gamma_2$ tends towards a constant
non-vanishing value which is proportional to the square of the atom 
number $N$. A higher condensate density thus significantly increases the
damping of the collective excitation.


\section{BEC in a ring cavity}

In this section we will now discuss the case of a BEC in a ring cavity. In this
case the condensate is coupled to the two independent travelling wave modes
$\alpha_{\pm}$. For simplicity we will assume in the following that both modes
of the cavity are driven with the same pumping rate $\eta$ \cite{Moore}. 
Therefore the equations of motion read
\begin{mathletters}
\label{eq:tr}
\bea
\frac{d}{dt} \alpha_{\pm}(t) & = &
   \left[ i\Delta_c - i N U_0 -\kappa \right]\alpha_{\pm}(t) \nonumber \\
 & &   -i N U_0 \langle e^{\mp 2ik\hat{x}}\rangle \alpha_{\mp} + \eta,
   \label{eq:atr} \\
i \frac{d}{dt} \psi(x,t) & = &
 \Bigl\{ \frac{\hat{p}^2}{2m} + Ng_{coll}|\psi(x,t)|^2 \nonumber \\
 & &  + U_0\left| \alpha_+(t) e^{ikx} + \alpha_-(t) e^{-ikx} \right|^2
     \Bigr\} \nonumber \\
 & & \times \psi(x,t). \label{eq:GPEtr}
\eea
\end{mathletters}
Hence in general the condensate will scatter light between the left and right
running waves and induce a strong coupling. This gives additional degrees of
freedom to the system compared to the standing wave case. For example, in
addition to intensity shifts the condensate can also induce a relative phase
shift between the two modes, which changes the position of the potential wells. 
Analogously, the minima can also be controlled externally by the relative
phase of the two driving fields which allows to selectively excite antisymmetric
excitations.

Considering the important role which the spatial symmetry plays for the standing
wave cavity, we will now change to the description of the cavity modes by
$\alpha_s=\alpha_++\alpha_-$ and $\alpha_a=\alpha_+-\alpha_-$ which have
symmetric and antisymmetric mode functions, respectively. In this new basis
Eqs.~(\ref{eq:tr}) read
\begin{mathletters}
\label{eq:trs}
\bea
\frac{d}{dt} \alpha_s(t) & = &
   \left[ i\Delta_c - i N U_0 -i N U_0 \langle \cos(2k\hat{x})\rangle
   - \kappa \right]\alpha_s(t) \nonumber \\
 & &   + N U_0 \langle \sin(2k\hat{x})\rangle \alpha_a(t) + 2\eta,
   \label{eq:astrs} \\
\frac{d}{dt} \alpha_a(t) & = &
   \left[ i\Delta_c - i N U_0 +i N U_0 \langle \cos(2k\hat{x})\rangle
   - \kappa \right]\alpha_a(t) \nonumber \\
 & &   - N U_0 \langle \sin(2k\hat{x})\rangle \alpha_s(t),
   \label{eq:aatrs} \\
i \frac{d}{dt} \psi(x,t) & = &
 \Bigl\{ \frac{\hat{p}^2}{2m} + Ng_{coll}|\psi(x,t)|^2 \nonumber \\
 & & + U_0\left| \alpha_s(t) \cos(kx) + i\alpha_a(t) \sin(kx) \right|^2
     \Bigr\} \nonumber \\
 & & \times \psi(x,t). \label{eq:GPEtrs}
\eea
\end{mathletters}
Note that because of the assumption of a single pumping rate $\eta$ for
$\alpha_+$ and $\alpha_-$, in the new basis only the symmetric mode $\alpha_s$
is pumped. The antisymmetric mode $\alpha_a$ only contains photons which have
been scattered by the condensate out of $\alpha_s$.

\subsection{Ground state}

For the calculation of the ground state of the compound system formed by the BEC
and the cavity modes we will again assume the case $U_0>0$. We then find that
the ground state wave function is localized at the antinodes of the driven mode
$\alpha_s$ and is symmetric in $x$. Thus the expectation values of $\sin(2kx)$
in Eqs.\ (\ref{eq:trs}) vanish and Eq.\ (\ref{eq:aatrs}) decouples. The
stationary state of the antisymmetric mode is therefore given by
$\alpha_{a,0}=0$. Equations (\ref{eq:astrs}) and (\ref{eq:GPEtrs}) then reduce
to the equations (\ref{eq:st}) for the standing wave cavity if one identifies
the parameters
\bea
2 U_0^{\text{r}} & = & U_0^{\text{s}}, \nonumber \\
|\alpha_s^{\text{r}}|^2/2 & = & |\alpha^{\text{s}}|^2, \nonumber \\
\sqrt{2} \eta^{\text{r}} & = & \eta^{\text{s}},
\label{eq:id}
\eea
for the ring cavity and the standing wave cavity, respectively. The ground state
of the system can thus be obtained by using our previous results for the
standing wave cavity and all of the discussions there equally apply to the
ground state in the ring cavity.

\subsection{Collective excitations}

The collective excitations are calculated with the same method as in the
preceding section by linearization of the equations of motion (\ref{eq:trs}) in
small deviations of $|\psi\rangle$, $\alpha_s$ and $\alpha_a$ from their
stationary states $|\psi_0\rangle$, $\alpha_{s,0}$ and $0$. Choosing the ground
state wave function to be real and taking its symmetry into account we obtain
\begin{mathletters}
\label{eq:trlin}
\bea
i \frac{d}{dt}\delta\alpha_s & = & 
   [-\Delta_c+2NU_0\langle \psi_0|\cos^2(k\hat{x})|\psi_0\rangle -i\kappa]
       \delta\alpha_s \nonumber \\
 & & + 2NU_0\alpha_{s,0} [\langle\delta\psi|\cos^2(k\hat{x})|\psi_0\rangle
   + c.c.], 
   \label{eq:trlinas} \\
i \frac{d}{dt}\delta\alpha_a & = & 
   [-\Delta_c+2NU_0\langle \psi_0|\sin^2(k\hat{x})|\psi_0\rangle -i\kappa]
       \delta\alpha_a \nonumber \\
 & & - NU_0\alpha_{s,0} [\langle\delta\psi|\sin(2k\hat{x})|\psi_0\rangle
   + c.c.], 
   \label{eq:trlinaa} \\
i \frac{d}{dt}\delta\psi & = & 
   \Bigl\{ \frac{\hat{p}^2}{2m} + |\alpha_{s,0}|^2 U_0 \cos^2(k\hat{x}) + 
      2Ng_{coll}|\psi_0|^2 \nonumber \\
 & &     - \mu\Bigr\} \delta\psi
     + Ng_{coll}\psi_0^2 \delta\psi^* \nonumber \\
 & & + U_0 \cos^2(k\hat{x})\psi_0 (\alpha_{s,0}\delta\alpha_s^* + c.c.)
   \nonumber \\
& & - \case{i}{2}U_0 \sin(2k\hat{x})\psi_0 (\alpha_{s,0}\delta\alpha_a^* -c.c.).
\label{eq:trlinpsi}
\eea
\end{mathletters}
From these equations we see that the behaviour of the excitation eigenstates
strongly depends on their spatial symmetry.

For {\em symmetric} excitations $\delta\psi(x)$ the last expectation value in
(\ref{eq:trlinaa}) vanishes and the antisymmetric cavity mode decouples from
the wave function. Hence in this case we find $\delta\alpha_a=0$. The equations
of motion for $\delta\psi$ and $\delta\alpha_s$ then reduce to their standing
wave counterpart discussed in the previous section if one rescales the
parameters as in (\ref{eq:id}). The symmetric collective excitations are thus
the same as those in a standing wave cavity.

Analogously, for {\em antisymmetric} excitations $\delta\psi(x)$ the symmetric
cavity mode decouples and therefore $\delta\alpha_s=0$. We then find a new set
of coupled equations for $\delta\psi$ and $\delta\alpha_a$. Thus, in contrast to
the case of a standing wave cavity, also the antisymmetric excitations are
damped in a ring cavity. However, the damping mechanism is of completely
different physical origin. Instead of the Sisyphus mechanism discussed above,
here the coherent scattering of photons from the $\alpha_s$ cavity mode into
the $\alpha_a$ mode is responsible for the damping. This leads to less severe
requirements for the cavity parameters as we will see in the following
subsection.

\begin{figure}[tb]
\infig{22em}{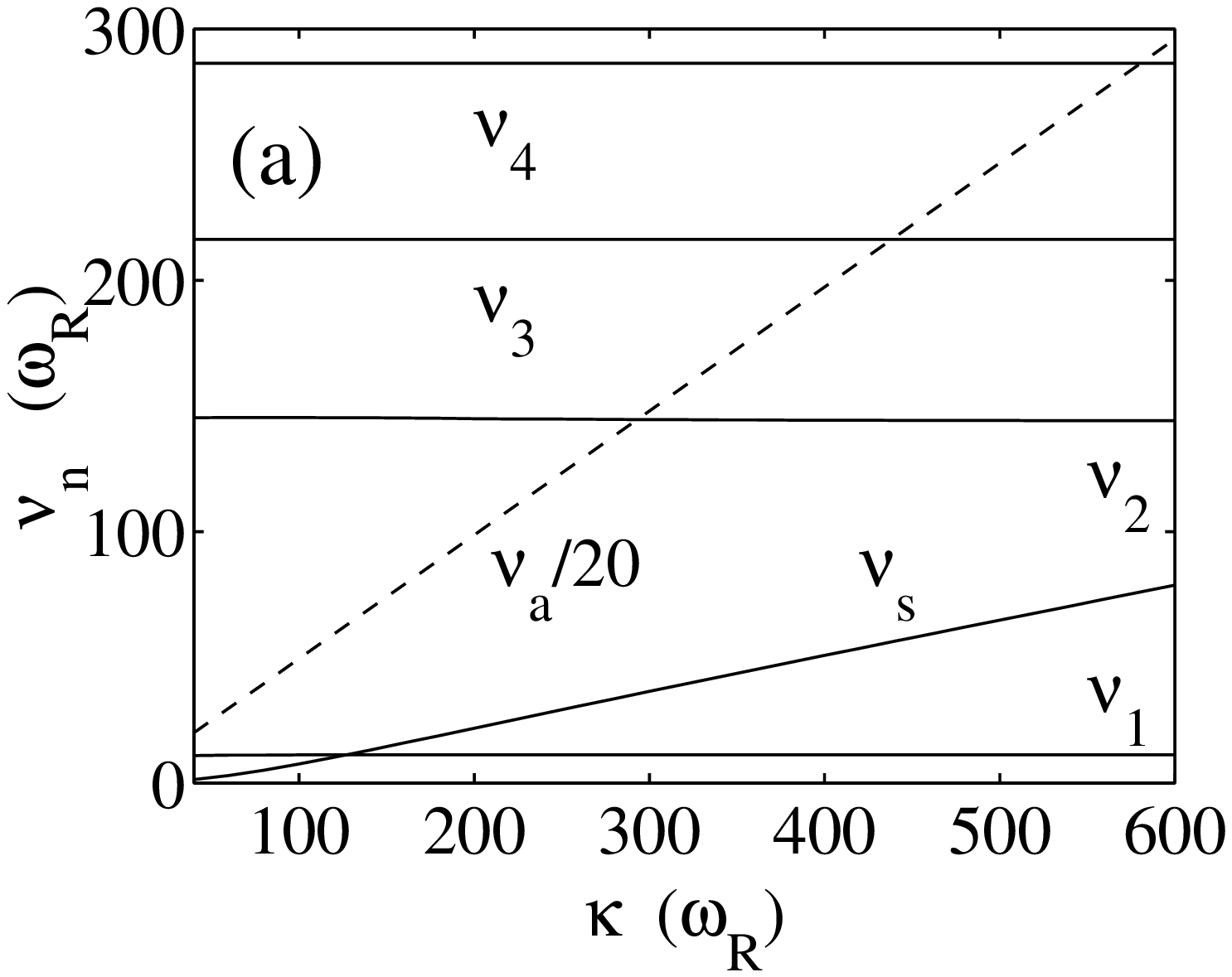}
\infig{22em}{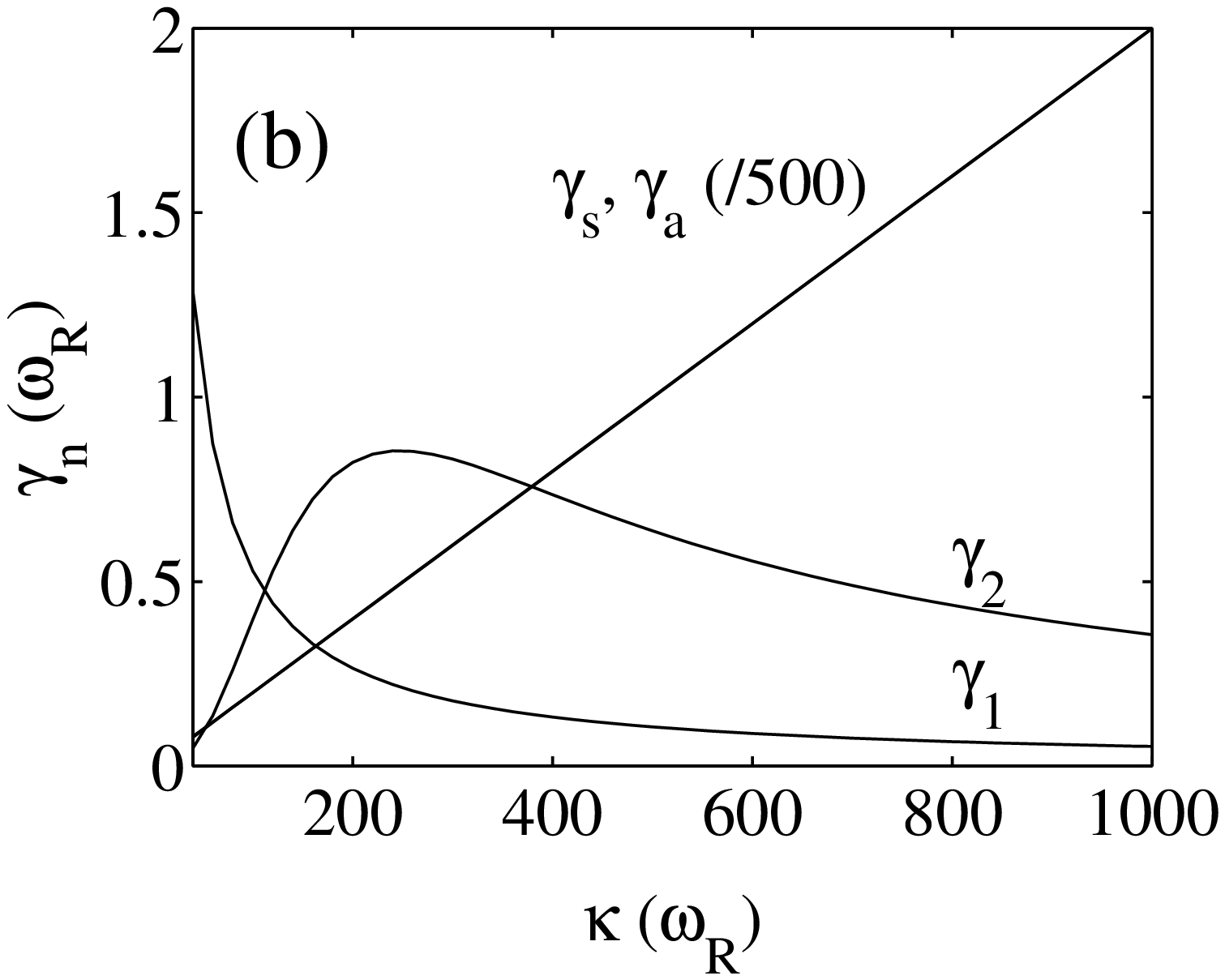}
\caption{Collective condensate excitations in a ring cavity: 
(a) frequencies $\nu_n$ and (b) 
 decay rates $\gamma_n$ vs cavity decay rate $\kappa$. The parameters are
 $g_{coll}=\Delta_c=0$, $N U_0=5\kappa$, $\eta^2=70\kappa N \omega_R$.
}
\label{fig:B}
\end{figure}

Figure \ref{fig:B} shows the spectrum of collective excitations of a Bose
condensate in a ring cavity which is obtained from the numerical solution of
Eqs.~(\ref{eq:trlin}). First, we note that in contrast to the case of the
standing wave cavity we now find {\em two} modes with eigenvalues which scale
proportional to the cavity decay rate $\kappa$. In the semiclassical limit
(atoms well localized), these correspond to pure oscillations of the symmetric
and antisymmetric field mode, respectively, and are thus labelled
$\omega_s=\nu_s-i\gamma_s$ and $\omega_a=\nu_a-i\gamma_a$. The semiclassical
limits of these eigenfrequencies are obtained from Eqs.~(\ref{eq:trlin}) as
$\omega_s=-\Delta_c-i\kappa$ and $\omega_a=-\Delta_c+2N U_0-i\kappa$. Although
the damping rates of these modes are equal, we see that the different spatially
depending coupling to the atoms leads to a large difference in the oscillation
frequencies.

For the parameters chosen in Fig.~\ref{fig:B} the other oscillation frequencies
$\nu_n$, $n\ge 1$, are mainly independent of $\kappa$. However, whereas all
frequencies with $n\ge 2$ are equally spaced and hence very well described by
harmonic oscillator states, the lowest frequency $\nu_1$ is significantly
shifted downwards. For the damping rates we find that only the two lowest
excitations exhibit relevant damping. However, the dependence of these damping
rates on the system parameters is qualitatively different according to the
different damping mechanisms. We will return to the discussion of these features
in the following subsection where we derive analytic approximations for the
eigenvalues.

\subsection{Harmonic oscillator approximation}

Let us now calculate analytic estimates for the lowest oscillation frequencies
and damping rates along the lines of Sec.~\ref{sec:stharm}. We will thus 
again assume $\Delta_c=g_{coll}=0$.

As we have already seen, the calculation of the ground state and of the
symmetric collective excitations can be reduced to the problem of the standing
wave if the appropriate identification of the system parameters (\ref{eq:id}) 
is made. We can therefore use our previous results to obtain the self-consistent
cavity field
\be
|\alpha_{s,0}|^2 = \frac{4\eta^2}{\kappa^2} - 
    \frac{N^2 U_0 \omega_R}{\kappa^2}
\ee
and the corresponding harmonic oscillator frequency
\be
\omega_0 = 2 \omega_R \sqrt{|\alpha_{s,0}|^2 U_0/\omega_R}.
\ee
For the lowest symmetric excitation, the expansion for large values of $\kappa$
yields 
\be
\nu_2 = 2\omega_0\sqrt{1-\frac{N^2 U_0 \omega_R}{4\eta^2}} 
\ee
and
\be
\gamma_2 = \frac{16 N^2 U_0^2 \omega_R^2}{\kappa^3}
   \left(1-\frac{N^2 U_0 \omega_R}{4\eta^2} \right)^2.
\ee

Analogously, we can calculate the lowest antisymmetric excitation by expanding
the expectation values in Eqs.~(\ref{eq:trlinaa}) and 
(\ref{eq:trlinpsi}) to lowest order in $k\hat{x}$. To this order only
$\delta\psi$ and $\delta\psi^*$ proportional to the first harmonic oscillator
wave function $\psi_1$ couple to the cavity field $\delta\alpha_a$ and
$\delta\alpha_a^*$ and we thus have to find the eigenvalues of a 4x4 matrix,
that is, we must solve the characteristic polynomial
\bea
& & [-i\kappa+2NU_0(1-\case{\omega_R}{\omega_0})-\omega]
[-i\kappa-2NU_0(1-\case{\omega_R}{\omega_0})-\omega]\nonumber \\
& & \times (\omega^2-\omega_0^2) 
    -4(NU_0\omega_0)^2(1-\case{\omega_R}{\omega_0})=0.
\label{eq:poltr}
\eea
In the limit of $\kappa\rightarrow\infty$ (with constant
$U_0/\kappa$ and $\eta^2/\kappa$) this yields the oscillation
frequency
\be
\nu_1 = \omega_0\sqrt{1-\frac{4N^2U_0^2(1-\case{\omega_R}{\omega_0})}
                        {\kappa^2+4N^2U_0^2(1-\case{\omega_R}{\omega_0})^2}}.
\label{eq:nu1tr}
\ee
The first order correction in $1/\kappa$ gives the dominant term of the
corresponding damping rate
\be
\gamma_1 = \omega_0^2\kappa\frac{4N^2U_0^2(1-\case{\omega_R}{\omega_0})}
           {[\kappa^2+4N^2U_0^2(1-\case{\omega_R}{\omega_0})^2]^2}.
\label{eq:g1tr}
\ee
We can now compare the behaviour of the two lowest eigenvalues as a function of
the system parameters. As an example, let us consider the case of a relatively
strong pump, $\eta^2\gg N^2U_0\omega_R$. In this limit, the second excitation
frequency $\nu_2$ is only weakly shifted from the harmonic oscillator frequency
$2\omega_0$. On the other hand, the frequency shift of the lowest excitation
$\nu_1$ mainly depends on the ratio $NU_0/\kappa$. Since this ratio has to be
larger than one in order to yield a significant frequency shift of the cavity by
the atoms, equation (\ref{eq:nu1tr}) implies that $\nu_1$ is strongly shifted
towards zero. Simultaneously we find for the damping rates that $\gamma_2$
becomes independent of $\eta$ in this limit, in contrast to $\gamma_1$ which is
proportional to $\omega_0$ and thus proportional to $\eta^2$. Therefore the
damping rate of the first antisymmetric excitation can be increased arbitrarily
by increasing the intensity of the pump field. The damping rate of the first
symmetric excitation is much harder to manipulate because it is mainly governed
by the optical potential per photon and thus by the quality of the cavity. On
the other hand, we note that $\gamma_2$ scales proportional to $N^2$ whereas 
$\gamma_1$ is inversely proportional to $N^2$. The number of atoms thus provides
another handle to change the relative size of the damping rates $\gamma_1$ and
$\gamma_2$.

Another point is worth a comment here. We emphasized in the previous section
that the damping mechanism for the collective excitations in a standing wave
cavity is crucially related to the width of the matter wave function and
vanishes in the semiclassical limit where the atoms are treated as point
particles. In contrast to this we find that in the travelling wave cavity the
damping mechanism still exists in the semiclassical limit. In fact, our results
for the oscillation frequency (\ref{eq:nu1tr}) and the damping rate
(\ref{eq:g1tr}) agree with the semiclassical results \cite{Gangl} if one 
takes formally the limit $\omega_R/\omega_0\rightarrow 0$.

\begin{figure}[tb]
\infig{22em}{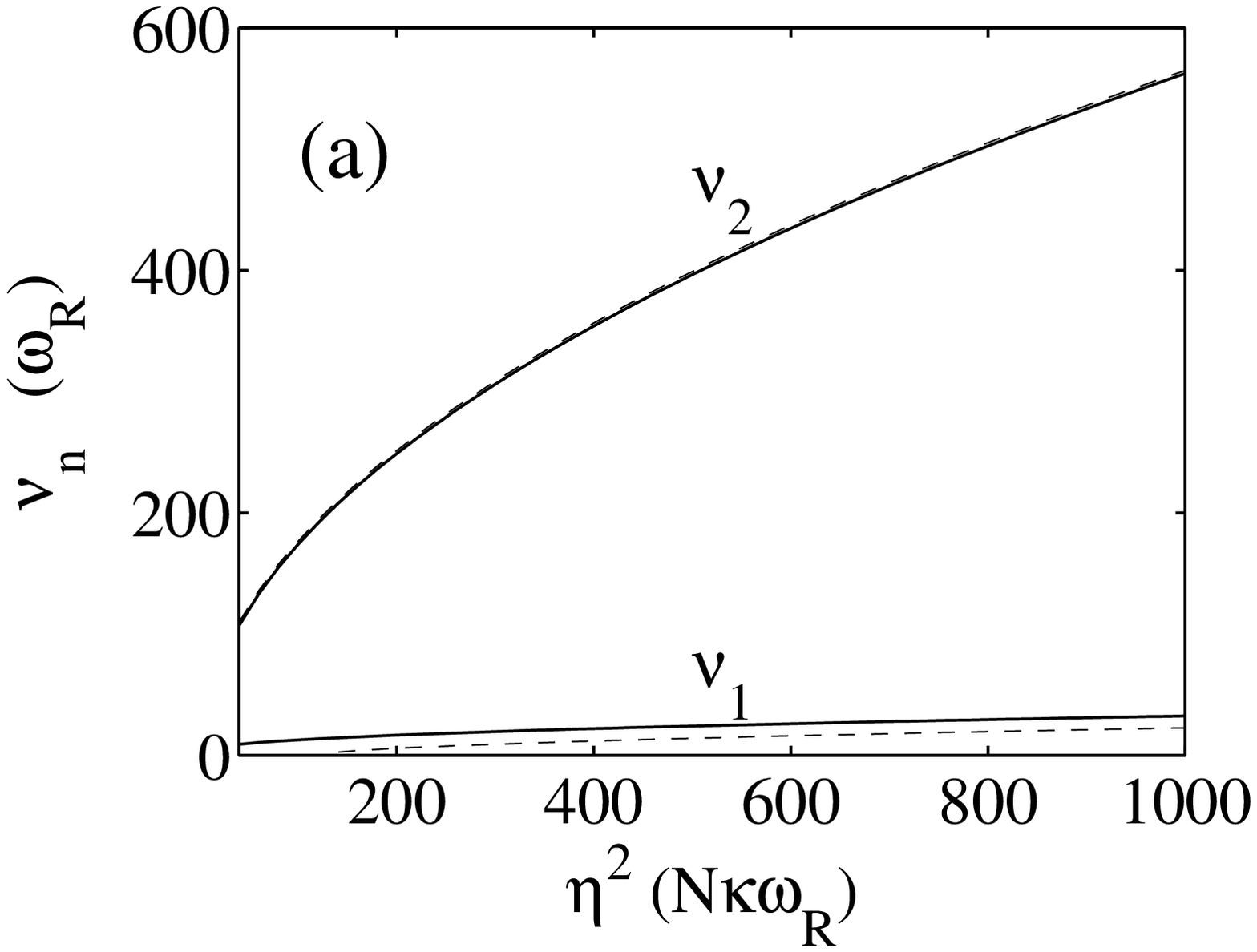}
\infig{22em}{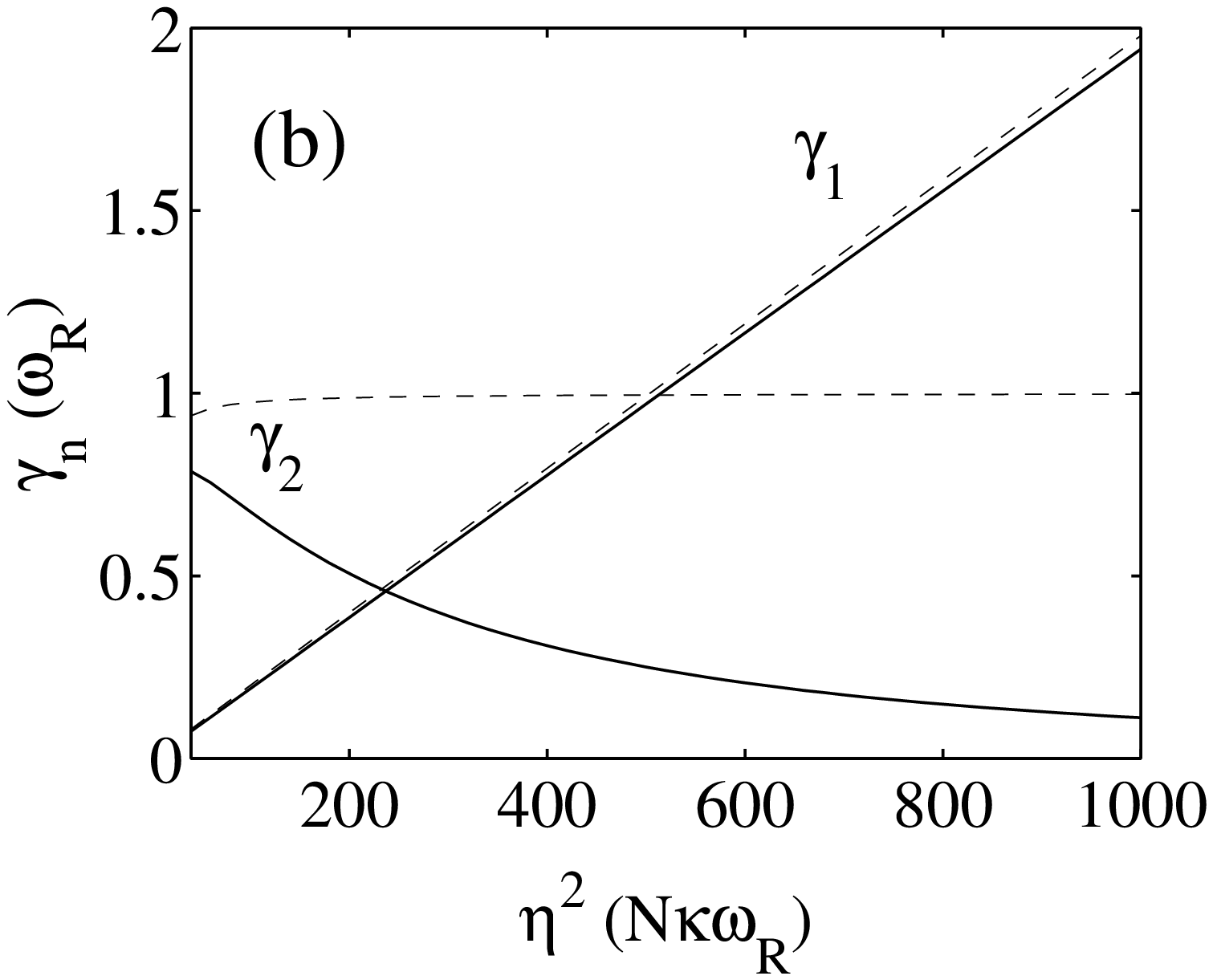}
\caption{Collective condensate excitations in a ring cavity: 
(a) frequencies $\nu_n$ and (b) 
decay rates $\gamma_n$ vs pump strength $\eta^2$. The parameters are
$g_{coll}=\Delta_c=0$, $N U_0=5\kappa$, $\kappa=400\omega_R$. The solid curves
are obtained numerically, the dashed curves are the analytical solutions
obtained in the harmonic oscillator approximation.
}
\label{fig:C}
\end{figure}

In Fig.~\ref{fig:C} we show the excitation frequencies $\nu_{1,2}$ and the
damping rates $\gamma_{1,2}$ as a function of the pump strength $\eta^2$ for
both the numerical solution and the analytic approximations. We see
that for the chosen parameters the approximations fit quite well apart from the
values of $\gamma_2$. This comes from the fact that we obtained the complex
eigenvalues $\omega_n$ from an expansion of
Eqs.~(\ref{eq:polst}) and (\ref{eq:poltr}) 
for small values of $|\omega_n|/\kappa\ll 1$. As we see from Fig.~\ref{fig:C}(a)
this is well fulfilled for $\omega_1$ for the chosen parameters, but 
$|\omega_2|/\kappa$ is of the order of one. However, in the limit of a strong
pump the lowest order term for the frequency $\nu_2$ already gives the correct
value, namely twice the harmonic oscillator frequency. Thus, only the imaginary
part (the damping rate $\gamma_2$) of the analytic approximation deviates from
the exact solution in Fig.~\ref{fig:C}. In parameter regions where
$|\omega_2|/\kappa\ll 1$ we find a much better agreement of the two solutions.

\section{Interacting Bose gas}

In the discussion so far we have omitted the effects of atomic interactions,
as described by
the collision rate $g_{coll}$ in the GPE, on the energies and damping rates of
the collective excitations. Neglecting this has allowed us to obtain analytical
expressions and therefore to discuss the parameter dependence of our results
explicitly. However, atomic collisions are known to play a crucial role in
experimental realizations of Bose-Einstein condensates. We will now discuss the
changes of the collective excitations according to collisions in a numerical
example of a condensate in a ring cavity.

\begin{figure}[tb]
\infig{22em}{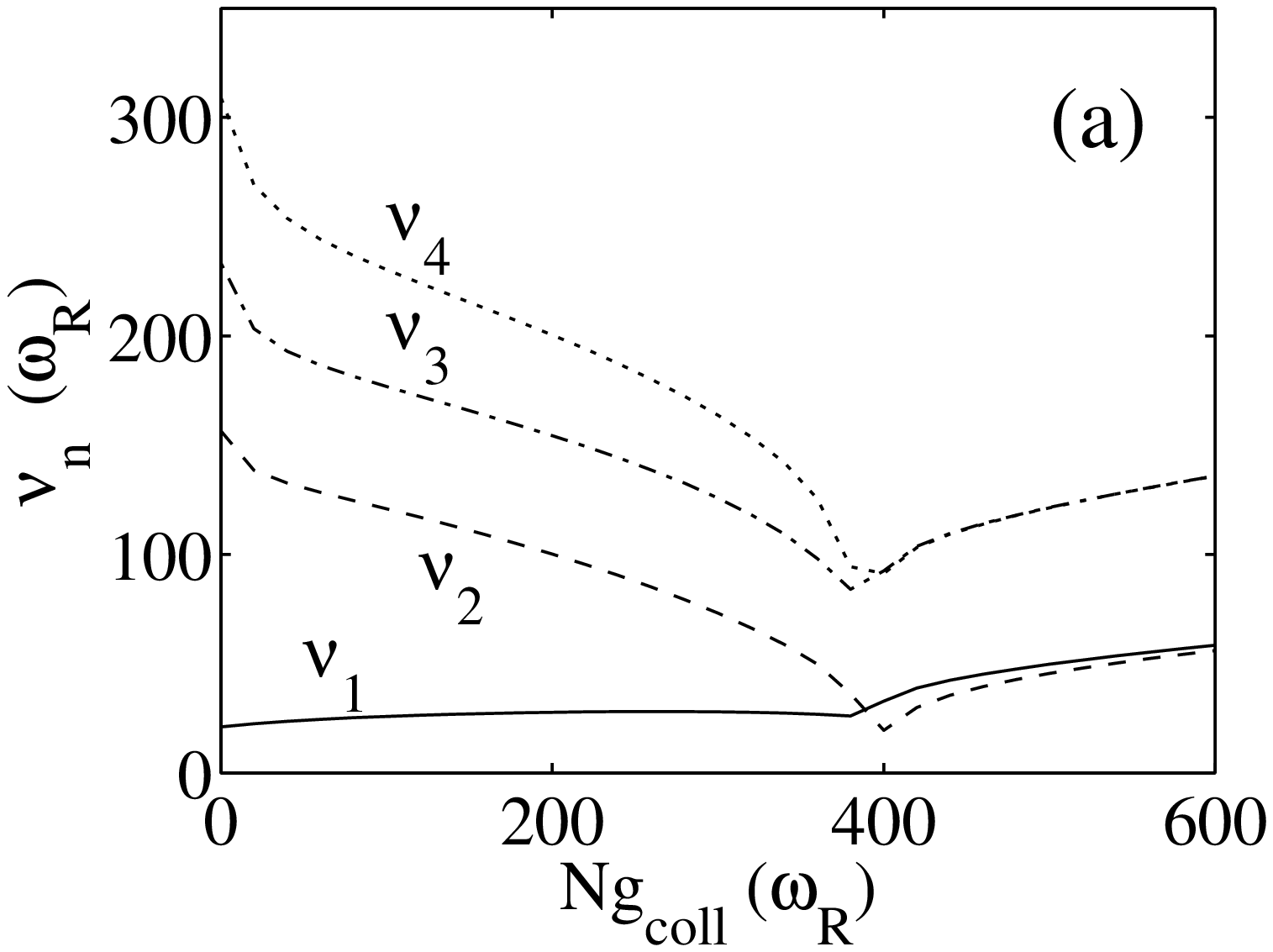}
\infig{22em}{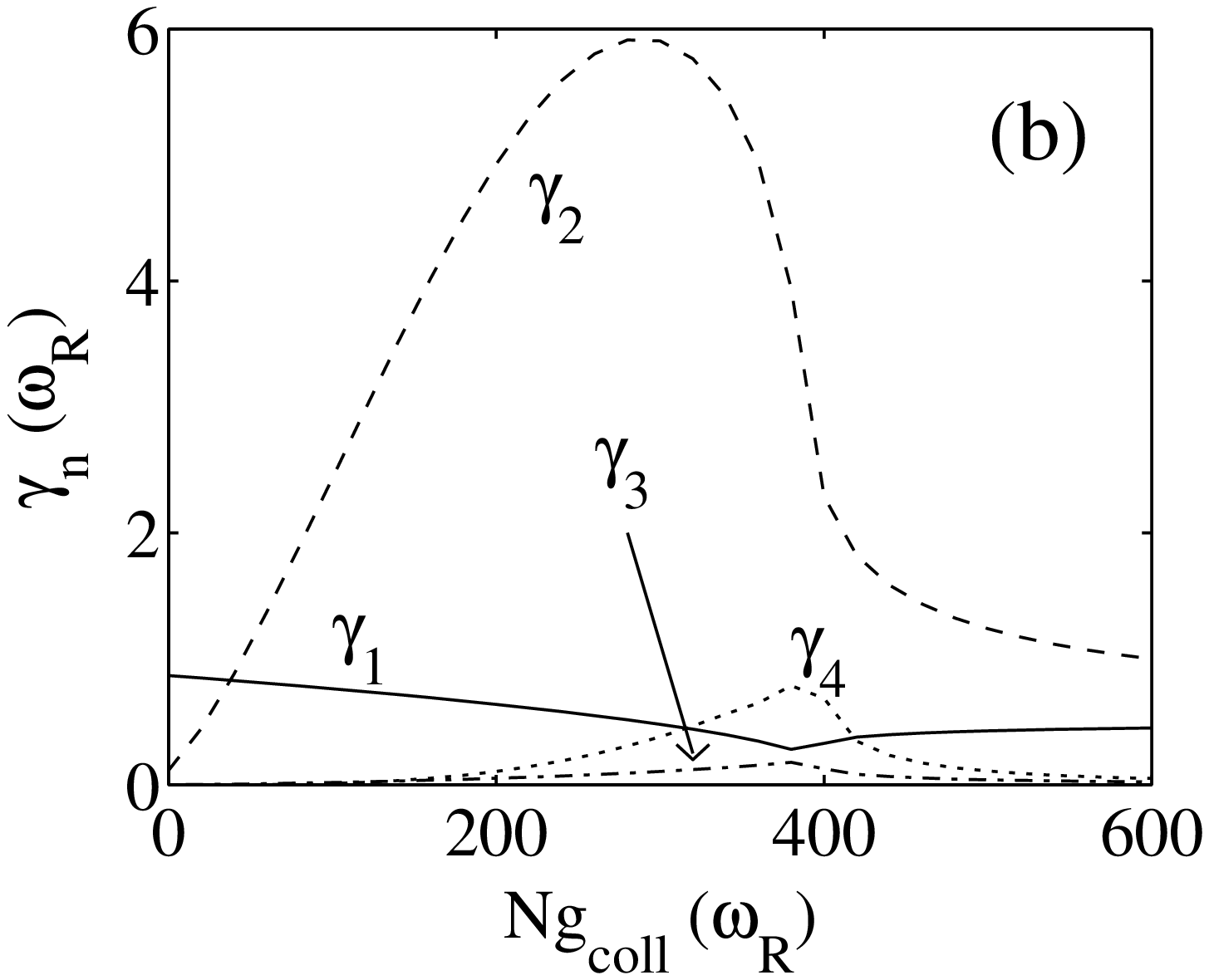}
\caption{(a) Frequencies $\nu_n$ and (b) 
decay rates $\gamma_n$ of the collective excitations vs atomic collision
rate $g_{coll}$. The parameters are
$g_{coll}=\Delta_c=0$, $N U_0=2\kappa$, $\kappa=400\omega_R$,
$\eta^2=200N\kappa\omega_R$.
}
\label{fig:D}
\end{figure}

We show in Fig.~\ref{fig:D} the excitation frequencies $\nu_n$ and the
corresponding damping rates $\gamma_n$ as a function of the collision rate
$g_{coll}$ with all other parameters fixed. The main effect on the 
stationary ground state wave function \cite{PRA}
of a repulsive interaction between the condensed atoms
is to increase the width of the wave function. Since this larger width also
changes the coupling to the cavity field, we find that the steady state photon
number decreases with increasing collision rate. Consequently, the optical
potential becomes more shallow and the excitation frequencies decrease. However,
as we can see from Fig.~\ref{fig:D} this argument thus not hold for the lowest
(antisymmetric) excitation. Here the atomic collisions counteract the strong
frequency shift which we found in the previous section and $\nu_1$ slightly
increases with $g_{coll}$. Above a certain threshold value for $g_{coll}$ the
atom-atom repulsion gets stronger than the confining effect of the optical
potential. In this case the ground state wave function is no longer localized
and the spectrum of excitations changes into that of unbound particles where
each excitation frequency is doubly degenerate.

Figure \ref{fig:D}(b) shows that collisional effects have an even more important
influence on the damping rates of the collective excitations. We see that the
effect differs for the damping rates $\gamma_1$ and $\gamma_2$.
While $\gamma_1$ weakly decreases with increasing $g_{coll}$, $\gamma_2$
increases significantly over a broad range of values of $g_{coll}$. This is
related to the fact that the damping of the symmetric excitation $\gamma_2$
depends crucially on the width of the ground state wave function whereas the
damping mechanism of the antisymmetric excitation does not, as already
emphasized before. Since the major effect of atom collisions is to broaden the
wave function, the resulting changes of the damping rates occur predominantly
for the symmetric excitations. 
Note also that the atomic collisions and the stronger anharmonicity of the
potential according to the lower field intensity enhance the damping of higher
collective excitations, as can be seen from the damping rates 
$\gamma_3$ and $\gamma_4$ in Fig.~\ref{fig:D}(b).


\section{Conclusions}

In summary we have studied in detail the interaction of a Bose-Einstein
condensate with one or two single modes in a high-finesse optical cavity. We
have solved the coupled set of non-linear equations of motion for the joint
dynamics of the condensate and the light field numerically and compared it analytically with
a simplified model based on the Lamb-Dicke expansion. We find that, even without
atom-atom interaction, the oscillation frequencies are shifted with respect to
their values in a fixed external potential.

For a finite cavity response time the collective excitations are
damped or amplified depending on the cavity detuning, which can be easily
controlled externally. We identify two distinct mechanisms depending on the spatial
symmetry of the excitations. The damping mechanism in standing wave cavities
and for spatially symmetric excitations in ring cavities is due to a
Sisyphus type effect, which leads to larger cavity fields at times when the
condensate runs up potential hills than at times when the condensate runs down.
On average this effect extracts kinetic energy from the condensate which is
carried away by the cavity field. On the other hand, the damping mechanism for
the spatially antisymmetric excitations is only present in a ring cavity due to the scattering
of cavity photons between the two counterpropagating waves. This creates an intensity imbalance, 
which is counteracted by the cavity damping and hence leads to momentum dissipation.

The two damping mechanisms exhibit very distinct parameter dependences. Our analytical approximations show that in
the limit of strong cavity pumping the damping rate of the spatially symmetric
excitations becomes independent of the pump but scales proportional to the
square of the atom number $N^2$ while the damping rate of the antisymmetric
excitations is proportional to the pump field intensity and inversely
proportional to $N^2$, which implies less stringent requirements to cavity technology.

The difference between the damping/amplification rates of excitations with
different spatial symmetry could be used to manipulate a Bose-Einstein
condensate in a controlled fashion.  In addition, in a ring cavity setup we can
also excite oscillations by external phase and amplitude shifts of the pump
light. All the effects could of course be enhanced by tailored feedback of the
measured transmitted intensity onto the pump. This might give rise to useful
applications of such a system in the context of quantum information and quantum
computation in analogy to other recently proposed systems making use of
particles in optical lattices \cite{Jaksch2,Brennen,Hemmerich}.

\acknowledgments

We thank J.\ I.\ Cirac and P.\ Zoller for stimulating discussions.
This work was supported by the Austrian Science Foundation FWF (Project P13435).



\begin{thebibliography}{99}

\bibitem{bec0} M.\ H.\ Anderson, J.\ R.\ Ensher, M.\ R.\ Matthews, 
   C.\ E.\ Wieman, and E.\ A.\ Cornell, Science {\bf 269}, 198 (1995);
   C.\ C.\ Bradley, C.\ A.\ Sackett, J.\ J.\ Tollett, and
   R.\ G.\ Hulet, Phys.\ Rev.\ Lett.\ {\bf 75}, 1687 (1995);
   K.\ B.\ Davis, M.-O.\ Mewes, M.\ R.\ Andrews, N.\ J.\ 
   van Druten, D.\ S.\ Durfee, D.\ M.\ Kurn, and W.\ Ketterle, 
   Phys.\ Rev.\ Lett.\ {\bf 75}, 3969 (1995).

\bibitem{Ketterle} W.\ Ketterle, D.\ S.\ Durfee, and D.\ M.\ Stamper-Kurn, {\em
Making, probing and understanding Bose-Einstein condensates}, Proc.\ Int.\
School of Physics Enrico Fermi, edited by M.\ Inguscio, S.\ Stringari, and C.\
E.\ Wieman (at press); preprint cond-mat/9904034.

\bibitem{Walls} A.\ S.\ Parkins and D.\ F.\ Walls, Phys.\ Rep.\ {\bf 303}, 2 
(1998).

\bibitem{Stringari} F.\ Dalfovo, S.\ Giorgini, L.\ P.\ Pitaevskii, and S.\
Stringari, Rev.\ Mod.\ Phys.\ {\bf 71}, 463 (1999).

\bibitem{Phillips} E.\ W.\ Hagley, L.\ Deng, M.\ Kozuma, J.\ Wen, K.\ Helmerson,
S.\ L.\ Rolston, and W.\ D.\ Phillips, Science {\bf 283}, 1706 (1999);
M.\ Kozuma, L.\ Deng, E.\ W.\ Hagley, J.\ Wen, R.\ Lutwak, 
K.\ Helmerson, S.\ L.\ Rolston, and W.\ D.\ Phillips, 
Phys.\ Rev.\ Lett.\ {\bf 82}, 871 (1999).

\bibitem{Kasevich} B.\ P.\ Anderson and M.\ A.\ Kasevich, Science {\bf 282},
1686 (1998).

\bibitem{Choi} D.-I.\ Choi and Q.\ Niu, Phys.\ Rev.\ Lett.\ {\bf 82}, 
   2022 (1999).

\bibitem{Jaksch} D.\ Jaksch, C.\ Bruder, J.\ I.\ Cirac, C.\ W.\ Gardiner, 
   and P.\ Zoller, Phys.\ Rev.\ Lett.\ {\bf 81}, 3108 (1998).

\bibitem{Berg} K.\ Berg-S{\o}rensen and K.\ M{\o}lmer, Phys.\ Rev.\ A 
   {\bf 58}, 1480 (1998).

\bibitem{Marzlin} K.-P.\ Marzlin and W.\ Zhang, Phys.\ Rev.\ A {\bf 59}, 
   2982 (1999).

\bibitem{Hau} L.\ V.\ Hau, S.\ E.\ Harris, Z.\ Dutton, and C.\ H.\ Behroozi
   Nature {\bf 397}, 594 (1999).

\bibitem{superrad} S.\ Inouye, A.\ P.\ Chikkatur, D.\ M.\ Stamper-Kurn, J.\
Stenger, D.\ E.\ Pritchard, and W.\ Ketterle, Science {\bf 285}, 571 (1999); 
Appl.\ Phys.\ B {\bf 69}, 347 (1999).

\bibitem{Cornell} M.\ R.\ Matthews, B.\ P.\ Anderson, P.\ C.\ Haljan, D.\ S.\
Hall, C.\ E.\ Wieman, and E.\ A.\ Cornell, Phys.~Rev.~Lett.\ {\bf 83},
   2498 (1999).

\bibitem{Dalibard} K.\ W.\ Madison, F.\ Chevy, W.\ Wohlleben, and J.\ Dalibard,
   cond-mat/9912015.

\bibitem{Haroche} G.\ Nogues, A.\ Rauschenbeutel, S.\ Osnaghi, M.\ Brune, J.\ 
M.\ Raimond, and S.\ Haroche, Nature {\bf 400}, 239 (1999); A.\ Rauschenbeutel
{\em et al.}, Phys.\ Rev.\ Lett.\ {\bf 83}, 5166 (1999).

\bibitem{Rempe} P.\ M\"unstermann, T.\ Fischer, P.\ Maunz, P.\ W.\ H.\ Pinkse,
   and G.\ Rempe, Phys.~Rev.~Lett.\ {\bf 82}, 3791 (1999); G.\ Rempe {\em et
   al.} in ``Laser Spectroscopy IX'', R.\ Blatt (ed.), (World Scientific,
   Singapore, in press).

\bibitem{Kimble} J.\ Ye, D.\ W.\ Vernooy, and H.\ J.\ Kimble, Phys.\ Rev.\
Lett.\ {\bf 83}, 4987 (1999); 
   C.\ J.\ Hood, M.\ S.\ Chapman, T.\ W.\ Lynn, and H.\ J.\ Kimble, 
   Phys.\ Rev.\ Lett.\ {\bf 80}, 4157 (1998).

\bibitem{Milburn} J.\ F.\ Corney and G.\ J.\ Milburn, Phys.~Rev.~A {\bf 58},
   2399 (1998).

\bibitem{Law} C.\ K.\ Law and N.\ P.\ Bigelow, Phys.~Rev.~A {\bf 58}, 4791
   (1998).

\bibitem{Meystre} E.\ V.\ Goldstein, E.\ M.\ Wright, and P.\ Meystre, 
   Phys.\ Rev.\ A {\bf 57}, 1223 (1998);
   E.\ S.\ Lee {\em et al.}, Phys.\ Rev.\ A {\bf 60}, 4006 (1999).

\bibitem{cavcool1} P.\ Horak, G.\ Hechenblaikner, K.\ M.\ Gheri, H.\ Stecher, 
   and H.\ Ritsch, Phys.\ Rev.\ Lett.\ {\bf 79}, 4974 (1997); 
   G.\ Hechenblaikner, M.\ Gangl, P.\ Horak, and H.\ Ritsch, 
   Phys.\ Rev.\ A {\bf 58}, 3030 (1998).

\bibitem{cavcool2} M.\ Gangl and H.\ Ritsch, Phys.\ Rev.\ A {\bf 61}, 
011402 (2000); Eur.\ Phys.\ J.\ D {\bf 8}, 29 (2000); Phys.\ Rev.\ A {\bf 61},
(in press).

\bibitem{PRA} P.\ Horak, S.\ M.\ Barnett, and H.\ Ritsch, Phys.\ Rev.\ A 
{\bf 61}, 033609 (2000).

\bibitem{Stringari2} S.\ Stringari, Phys.\ Rev.\ Lett.\ {\bf 77}, 2360 (1996).

\bibitem{Perez} V.\ M.\ Perez-Garcia, H.\ Michinel, J.\ I.\ Cirac, M.\
Lewenstein, and P.\ Zoller, Phys.\ Rev.\ Lett.\ {\bf 77}, 5320 (1996).

\bibitem{You} L.\ You, W.\ Hoston, and M.\ Lewenstein, Phys.\ Rev.\ A {\bf 55},
R1581 (1997).

\bibitem{Castin} Y.\ Castin and R.\ Dum, Phys.\ Rev.\ A {\bf 57}, 3008 (1998).

\bibitem{Moore} M.\ G.\ Moore and P.\ Meystre, Phys.~Rev.~A {\bf 59}, R1754
   (1999); M.\ G.\ Moore, O.\ Zobay, and P.\ Meystre, Phys.~Rev.~A {\bf 60}, 
   1491 (1999).

\bibitem{Gangl} M.\ Gangl {\em et al.}, unpublished.

\bibitem{Jaksch2} D.\ Jaksch, H.-J\ Briegel, J.\ I.\ Cirac, C.\ W.\ Gardiner, 
   and P.\ Zoller, Phys.\ Rev.\ Lett.\ {\bf 82}, 1975 (1999).

\bibitem{Brennen} G.\ K.\ Brennen, C.\ M.\ Caves, P.\ S.\ Jessen, and I.\ H.\
Deutsch, Phys.\ Rev.\ Lett.\ {\bf 82}, 1060 (1999).

\bibitem{Hemmerich} A.\ Hemmerich, Phys.\ Rev.\ A {\bf 60}, 943 (1999).

\end{thebibliography}
\end{document}